\documentclass{emulateapj}
\usepackage{amsmath}
\usepackage{amssymb}
\usepackage{graphicx}
\usepackage{appendix}
\usepackage{color}
\graphicspath{{figs/}}
\def\gtaprx {\lower .1ex\hbox{\rlap{\raise .6ex\hbox{\hskip .3ex
        {\ifmmode{\scriptscriptstyle >}\else
                {$\scriptscriptstyle >$}\fi}}}
        \kern -.4ex{\ifmmode{\scriptscriptstyle \sim}\else
                {$\scriptscriptstyle\sim$}\fi}}}
\def\ltaprx {\lower .1ex\hbox{\rlap{\raise .6ex\hbox{\hskip .3ex
        {\ifmmode{\scriptscriptstyle <}\else
                {$\scriptscriptstyle <$}\fi}}}
        \kern -.4ex{\ifmmode{\scriptscriptstyle \sim}\else
                {$\scriptscriptstyle\sim$}\fi}}}
\newcommand{\lSect}[1]{{\label{sec:#1}}}
\newcommand{\Sectff}[1]{{\ref{sec:#1}}}
\newcommand{\Sect}[1]{{\S~\Sectff{#1}}}

\newcommand{\lFig}[1]{{\label{fig:#1}}}
\newcommand{\FIGFF}[2]{{\ref{fig:#2}{#1}}}

\newcommand{\FIG}[2]{{Fig.~\FIGFF{#1}{#2}}}
\newcommand{\Fig}[1]{{\FIG{}{#1}}}

\newcommand{\lEq}[1]{{\label{eq:#1}}}
\newcommand{\Eqref}[1]{{\ref{eq:#1}}}
\newcommand{\Eqff}[1]{{(\Eqref{#1})}}

\newcommand{\Eq}[1]{{eq.~\Eqff{#1}}}

\newcommand{\CASTRO}{\texttt{CASTRO}}
\newcommand{\KEPLER}{\texttt{KEPLER}}

\newcommand{\Ni}{{\ensuremath{^{56}\mathrm{Ni}}}}

\newcommand{\Ox}{{\ensuremath{^{16}\mathrm{O}}}}

\newcommand{\Cx}{{\ensuremath{^{12}\mathrm{C}}}}

\newcommand{\erg}{{\ensuremath{\mathrm{erg}}}}

\newcommand{\gcc}{\ensuremath{\mathrm{g}\,\mathrm{cm}^{-3}}}

\newcommand{\Msun}{\ensuremath{\mathrm{M}_\odot}}

\newcommand{\note}[1]{\emph{\textcolor{red}{}}}

\makeatletter

\newcommand{\Rmnum}[1]{\expandafter\@slowromancap\romannumeral #1@}
\makeatother

\begin{document}

\title{Magnetar-Powered Supernovae in Two Dimensions. I. Superluminous
  Supernovae}

\author{Ke-Jung Chen      \altaffilmark{1,2,3},
        S. E. Woosley      \altaffilmark{3}, and
        Tuguldur Sukhbold \altaffilmark{3}
        }

\altaffiltext{1}{Division of Theoretical Astronomy, 
                 National Astronomical Observatory of Japan, 
                 Tokyo 181-8588, Japan} 
\altaffiltext{2}{Institute of Astronomy and Astrophysics, 
                 Academia Sinica,  
                 Taipei 10617, Taiwan}
\altaffiltext{3}{Department of Astronomy \& Astrophysics, 
                 University of California, 
                 Santa Cruz, CA 95064, USA}

\altaffiltext{*}{EACOA Fellow, email: {\tt ken.chen@nao.ac.jp} } 

\begin{abstract}
Previous studies have shown that the radiation emitted by a rapidly
rotating magnetar embedded in a young supernova can greatly amplify
its luminosity.  These one-dimensional studies have also revealed the
existence of an instability arising from the piling up of radiatively
accelerated matter in a thin dense shell deep inside the
supernova. Here we examine the problem in two dimensions and find
that, while instabilities cause mixing and fracture this shell into
filamentary structures that reduce the density contrast, the
concentration of matter in a hollow shell persists. The extent of the
mixing depends upon the relative energy input by the magnetar and the
kinetic energy of the inner ejecta. The light curve and spectrum of
the resulting supernova will be appreciably altered, as will the
appearance of the supernova remnant, which will be shellular and
filamentary.  A similar pile up and mixing might characterize other
events where energy is input over an extended period by a centrally
concentrated source, e.g. a pulsar, radioactive decay, a
neutrino-powered wind, or colliding shells. The relevance of our
models to the recent luminous transient ASASSN-15lh is briefly
discussed.

\end{abstract}


%%%%%%%%%%%%%%%%%%%%%%%%%%%%%%%%%%%%%%%%%%%%%%%%
% Section 1 - Introduction                     %
%%%%%%%%%%%%%%%%%%%%%%%%%%%%%%%%%%%%%%%%%%%%%%%%
\section{INTRODUCTION} 
\lSect{intro}

Magnetars are neutron stars with unusually strong magnetic fields,
typically greater than 10$^{13}$ Gauss (G). Observational evidence
suggests that magnetars form in a significant fraction of supernovae
\citep{Kou98}, where the strong magnetic field may be a consequence of
the collapse of a rapidly differentially rotating iron core
\citep{Dun92,Tho93,Whe00,Tho04,Mos15}. It would thus not be surprising
if magnetars are also frequently born with rapid rotation rates which
they dissipate shortly after being born. Indeed, the so called
``millisecond magnetar'' is a popular model for the production of
long-soft gamma-ray bursts \citep[e..g.][]{Met11,Met15}.  
\citet{Maz14} and \citet{Can15} have pointed out that the upper bound of $2\times 10^{52}$ erg commonly 
	assumed for the most rapidly rotating neutron stars may be reflected in 
	an upper bound for the observed energy in supernovae accompanying GRBs.
There the required field strength approaches 10$^{16}$ G and the rotational
energy is about $2 \times 10^{52}$ erg or 20 Bethe (B), a substantial
fraction of which is emitted in 10 sec.  For less extreme field
strengths in the range 10$^{14}$ - 10$^{15}$ G and rotation periods
$\sim$ 5 ms, the assumption of pulsar-like emission implies that a
smaller amount of energy is emitted over a much longer time. Following
a suggestion by \citet{Mae07}, studies by \citet{Woo10} and
\citet{Kas10} showed that supernovae containing moderately energetic
magnetars can power exceptionally luminous transients sometimes
referred to as superluminous supernovae
\citep[e.g.][SLSNe]{Qui11,Gal12,Ins13}.  There, because of its late
time introduction, a substantial fraction of the total rotational
energy of the neutron star is emitted as light.

These same studies also revealed a shortcoming in the one-dimensional
models.  Since the rapidly rotating magnetar deposits an energy
comparable to the kinetic energy of the slower moving ejecta of the
original supernova, the deposition has consequences not only for the
brightness of the supernova, but for its dynamics as well. The
magnetar's energy, presumably initially in the form of x-rays or
gamma-rays and a wind, originates in a small volume. As a small amount
of matter carries a large amount of energy outwards, it ``snowplows''
into the overlying ejecta. In 1D studies, this causes a pile up of
most of the accelerated matter in a very thin shell. Eventually, the
density contrast between this shell and its surroundings, which can
approach a factor of 1,000 or more, causes numerical difficulty in the
simulation. If real, this pile up of most of the ejecta into a thin
shell would have consequences for the light curve and
spectrum. Radiation would be unrealistically trapped, at least
initially, inside the bubble it inflates, and the spectrum would show
a large amount of matter moving at just one speed. This is not a
realistic outcome.

Studies by \cite{Che82,Jun98,Blo01} have shown that similar thin
shells, formed by a pulsar wind in a supernova remnant, are
unstable. A similar instability might be expected to lead to the break
up of the shells in supernovae that magnetars accelerate.  Ideally, 3D
radiation-hydrodynamical simulations that well resolve both the energy
deposition region of the pulsar and the unstable thin shell would be
used to study this mixing and to obtain their light curves and
spectra.  Such simulations are beyond the present capability of our
numerical codes and computational resources.  As a first step, we have
carried out 2D hydrodynamical simulations using a realistic magnetar
progenitor, but neglecting radiation transport.  The neglect of
radiation transport is a reasonable approximation to the actual situation
since the density spike forms at an early phase when the matter is
still very optically thick and the radiation is strongly coupled with the
gas flow.

The supernova models studied here start from a 6 \Msun\ carbon-oxygen
(CO) core that has been previously evolved to the presupernova stage
\citep{Suk14}.  The star's 1.45 \Msun\ iron-core is assumed to
collapse to a magnetar. All external matter is ejected using a
piston so as to provide a final kinetic energy of 1.2 B. The source
of this initial explosion is unspecified, but could be either neutrino
transport or the action of a rapidly rotating magnetized proto-neutron
star itself. Any initial jet formation is neglected.

Two different magnetars are then embedded in these standrad ejecta,
both with a constant magnetic field strength of $4\times10^{14}$ G,
but having rotational energies either appreciably greater than or less
than the initial (1.2 B) explosion energy. The magnetar is assumed to
add power to the ejecta through its dipole emission. Its energy is
deposited in a small region, along within small amount of matter to
prevent the complete evacuation of the region that would result in it
becoming optically thin. At too low a density, the energy deposited
would also result in super-luminal motion since our code is not
relativistic.

The structure of the paper is as follows; in Section 2, the progenitor
model and the setup for the 2D simulations are described.  In Section
3 and 4, the results of the 2D simulations are given, and the
mechanics behind the formation of fluid instabilities discussed.  We
conclude in Section 5 and discuss the relevance of our 2D model for
the extreme case of a 1 ms magnetar embedded in a 6 \Msun \ core. This
might be relevant to the recently discovered transient, SLSN
candidate, ASASSN-15lh \citep{Don15}, if it is a magnetar-powered
supernova.  Some recent studies \citep[e.g.]{Met16,Hol16} have
suggested that ASASSN-15lh might be a ``tidal-disruption event'' (TDE)
instead of a supernova, but the actual situation is unclear at this
time.

%%%%%%%%%%%%%%%%%%%%%%%%%%%%%%%%%%%%%%%%%%%%%%%%
% Section 2 - Numerical Method %
%%%%%%%%%%%%%%%%%%%%%%%%%%%%%%%%%%%%%%%%%%%%%%%%
\section{Numerical Method}
\lSect{num_method}

\subsection{Presupernova Star}
\lSect{presn}

The progenitor is a 6 \Msun\ CO star with an initial mass fractions of
$\Cx$=0.14 and $\Ox$ = 0.86, as might result from the evolution of a
non-rotating solar metallicity star with a zero-age main sequence mass
of $\sim$24 \Msun. This model, previously published by \citet{Suk14},
has been followed, using the \KEPLER\ code, through carbon, neon, oxygen
and silicon burning and iron core collpase. It is presumed to have lost
all its envelope and part of its helium core to a wind or a binary
companion. A bare CO core was employed both for its simplicity of
modeling on a 2D Eulerian grid and because many SLSNe have been
observed to be Type I. If the star were a red or blue supergiant (RSG
or BSG), there would be additional mixing when the fractured CO core
ran into its low density hydrogen envelope.

The evolution of the CO core was followed until the collapse speed in
its iron core (1.45 \Msun) exceeded 1,000 km s$^{-1}$. The iron core
was then replaced with a gravitational point source and a parametrized
piston that moved so as to eject all matter external to the iron core
with a final kinetic energy, at infinity, without magnetar energy
deposition, of 1.2 B.  The explosion synthesized and ejected 0.22
\Msun\ of \Ni. The structure of the ejecta 100 s after the launches a
shock wave is shown in \Fig{presn}. At this time the original
supernova shock has already exited the star and the supernova is
coasting nearly homologously. The final velocity profile is very
similar to that shown in the figure.

%
% Fig 1
\begin{figure}[h]
\centering
\includegraphics[width=\columnwidth]{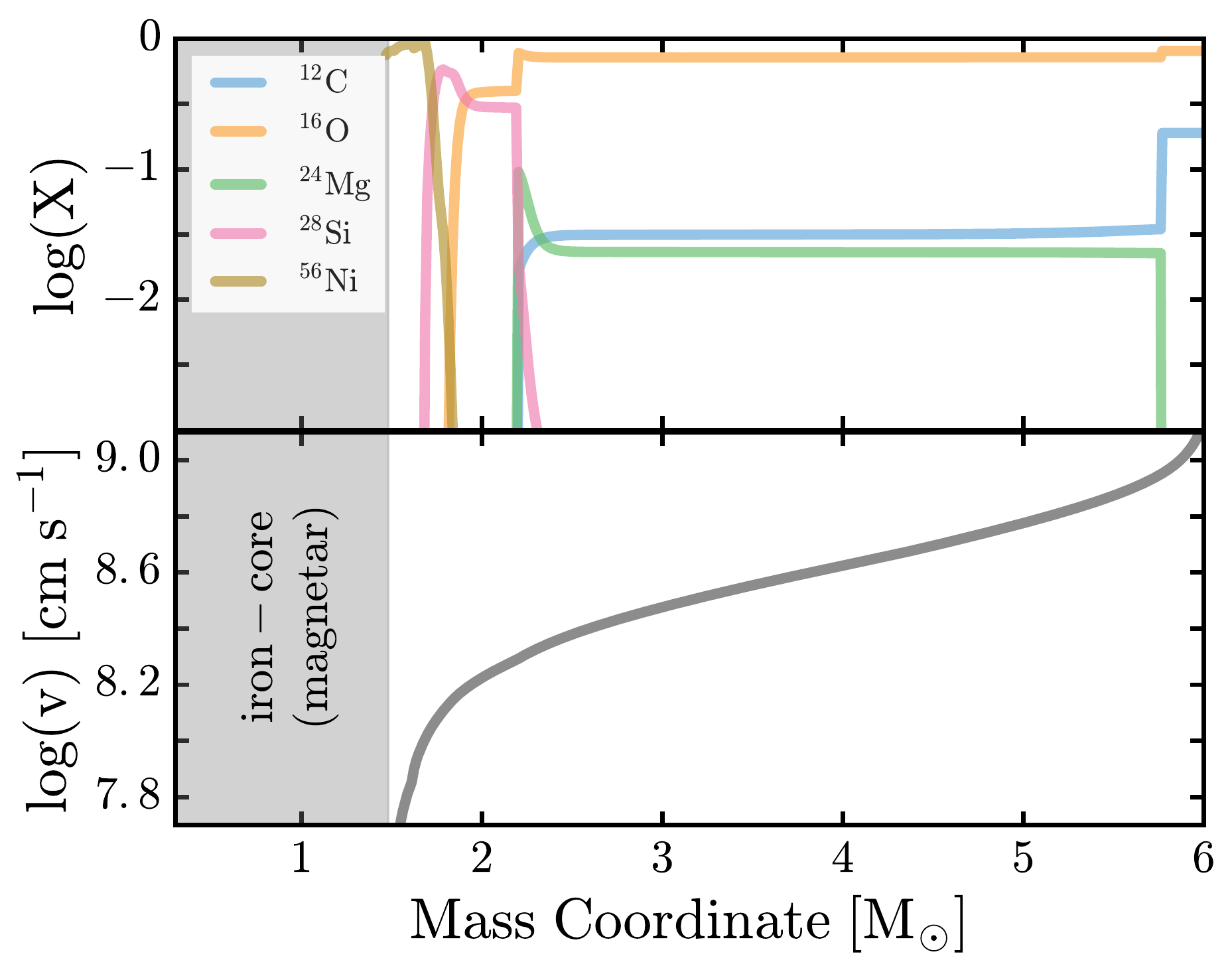} 
\caption{The starting model is a 6 \Msun\ CO core evolved to the
  presupernova stage by \citet{Suk14}. The iron core mass is 1.45
  \Msun (shaded gray). This core is assumed to collapse to a magnetar
  and eject all matter outside with a final kinetic energy of 1.2
  B. The mass fractions of selected isotopes (top) and the velocity
  profile (bottom) in the ejecta are shown prior to any energy
  deposition by the magnetar. The velocity structure shown is that of
  the supernova 100 s after core collapse at which point the initial
  shock wave has already passed through the surface of the
  presupernova star. The final velocity profile is nearly
  identical. 0.17 \Msun \ is ejected faster than $1 \times 10^{9}$ cm
  s$^{-1}$ and 0.0027 \Msun, faster than $2 \times 10^{9}$ cm
  s$^{-1}$. \lFig{presn}}
\end{figure}
\subsection{Magnetar Input}
\lSect{mag}

Starting 100 s after the initial explosion, a simulated magnetar
power source was introduced in the deep interior of the expanding
ejecta with a luminosity given by the Larmor formula \citep{Lyn90},
\begin{equation}
\begin{split}
\rm L_m & = -\frac{32\pi^4}{3c^2}\rm (BR_{ns}^3\sin\alpha)^2P^{-4}
\\ & \rm \approx -1.0 \times 10^{49}B_{15}^2 P_{ms}^{-4} \quad \erg\ s^{-1},
\end{split}
\lEq{magP}
\end{equation}
where the surface dipole field strength, $\rm B_{15}= B/10^{15}$G
is measured at the equator and the initial magnetar spin period $\rm
P_{\rm ms}$ is expressed in milliseconds.  The radius of the neutron
star is assumed to be $\rm R_{ns} = 10^{6}$ cm, and $\alpha$ is
the inclination angle between the magnetic and rotational axes, taken
$\rm \alpha = 30^{\circ}$.  Similar to \cite{Woo10}, the moment of
inertia for the neutron star is taken to be $\rm I = 10^{45}$ g
cm$^2$, thus the rotational kinetic energy is:
\begin{equation}
\rm E = \frac{1}{2}I\omega^2\approx 2\times 10^{52}P_{ms}^{-2} \quad
\erg.
\lEq{magE}
\end{equation}

It is common practice to take a limit of about 20 B and one
millisecond for the maximally rotating magnetar, though \citet{Met15}
have suggested a maximal value of 100 B in extreme cases. Assuming a
constant magnetic field, \Eq{magP} and \Eq{magE} imply that the magnetar
period, luminosity and energy evolution are given by
\begin{equation}
\begin{split}
\rm P(t) & \rm \approx (1+t/t_m)^{1/2}P_0\  ms,\\ 
\rm L(t) & \rm \approx (1+t/t_m)^{-2}E_0t_m^{-1}\ erg\ s^{-1},\\
\rm E(t) & \rm \approx (1+t/t_m)^{-1}E_0\ erg,
\end{split}
\lEq{evo}
\end{equation}
where $\rm P_0 = P_{\rm ms}(0)$, $\rm E_0=E(P_0)$ and $\rm t_m \approx
2\times10^3 P_{ms}^2B_{15}^{-2}$ is the magnetar spin-down
timescale. In the 1D \KEPLER\ calculations, the magnetar energy
generation is spread uniformly through the inner ten Lagrangian zones
of total mass $\sim \rm 2.4\times10^{32}$ gm, with an approximately
constant energy generation rate per gram. In the 2D Eulerian grid
calculations, the power is deposited in a constant volume bounded by a
radius of about $5 \times 10^9$ cm (1 ms run) and $5 \times 10^{11}$
cm (5 ms run), both corresponding to $2 - 3\%$ of the initial radius
of ejecta at the time the calculation began.  Both volumes are fixed
and resolved in the 2D study by about 100 grid points and the short
Courant time step in this small volume restricted the time scale of
the calculation.  Although \CASTRO\ sub-cycles in time step for
refined zones, the simulations still required numerous steps to evolve
and that made them computationally expensive.

The evolution of the magnetar luminosity for a range of B-fields and
two initial rotational rates is shown in \Fig{power}. The energy
deposition history is shown in black. 
%
% Fig 2 
\begin{figure}[h]
\centering
\includegraphics[width=\columnwidth]{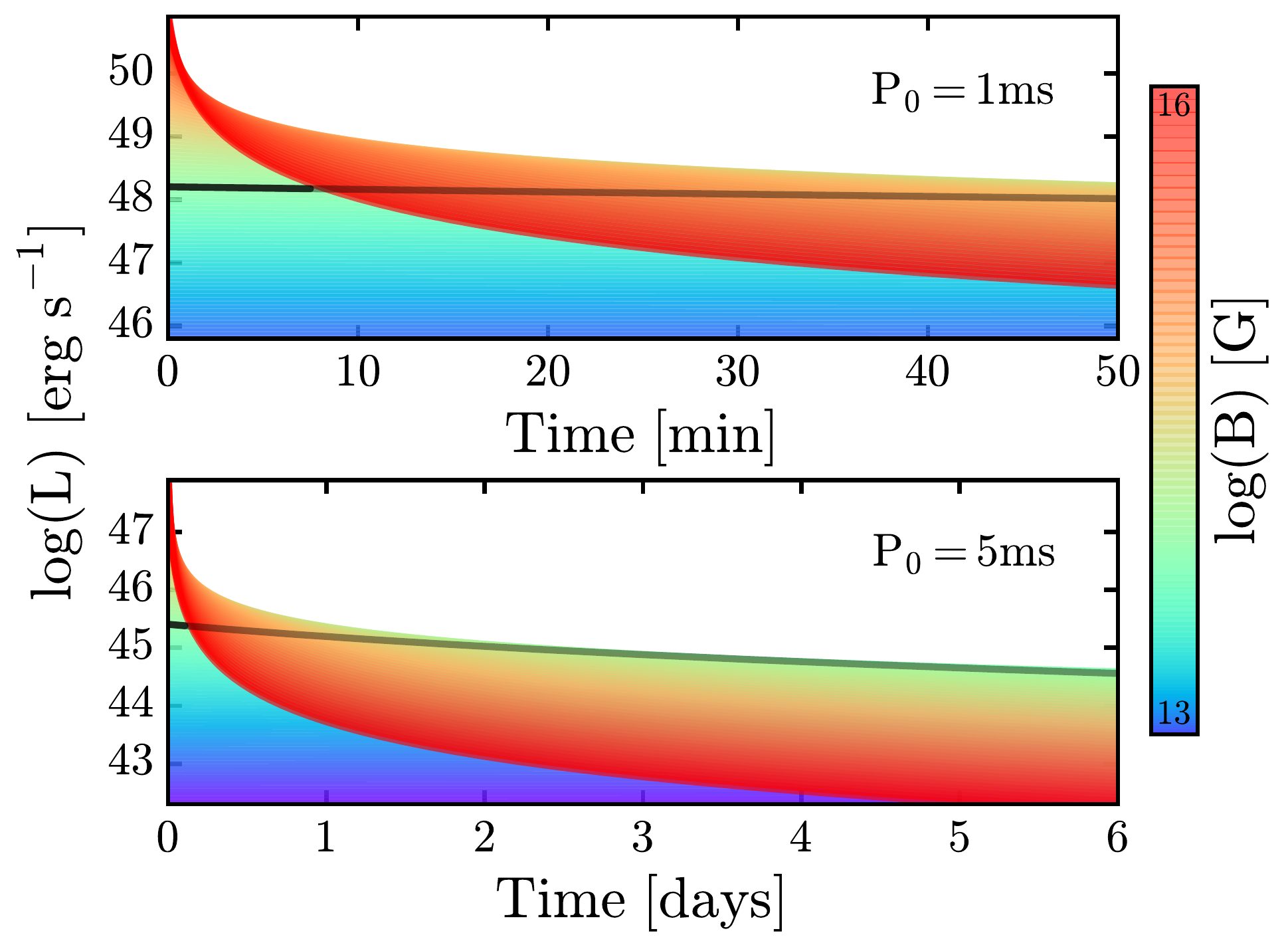} 
\caption{Energy deposition (eq. \Eqref{evo}) due to a 1 ms magnetar
  (top) and a 5 ms magnetar (bottom). Color coding indicates a range
  of possibilities depending upon the magnetic field strength which
  varies in the plot from $10^{13}$ to $10^{16}$ G.  The black lines
  correspond to a field strength of $4 \times 10^{14}$ G as was used
  in this paper and the time range corresponds to the duration of the
  2D studies in \Sect{results}. \lFig{power} }
\end{figure}

\subsection{2D \CASTRO{} Setup}
\lSect{castro}

\CASTRO\ is a multi-dimensional adaptive-mesh-refinement (AMR)
hydrodynamics code \citep{Alm10, Zha11}. It uses an unsplit
piecewise-parabolic method (PPM) hydro scheme \citep{Col84} with
multi-species advection and employs the Helmholtz equation of state
\citep{Tim00}, which includes electron and positron pairs of arbitrary
relativity and degeneracy, as well as ions, and radiation. Since the
density of the supernova ejecta is low $< 10^7 \gcc$, Coulomb
corrections to the equation of state were neglected.

In both the \KEPLER\ (1D) calculation and the \CASTRO\ (2D)
simulation, the effect of the magnetar was introduced 100 seconds
after the initial explosion. By this time, all the piston energy has
been deposited and the supernova has almost reached a coasting
configuration. During the neglected 100 seconds, the fiducial magnetar
with field strength $4\times10^{14}$ G would have deposited only
$0.6\%$ of its total rotational energy for the 1 ms case, and $0.032\%$
for the 5 ms case. This is small compared with either the total
rotational energy or initial explosion energy. If the magnetar played
a role in launching the explosion it must have had a larger field
strength at that time and used physics not encapsulated in the simple
dipole formula.  The time chosen for linking from \KEPLER\ to
\CASTRO\ was well before the development of any density spike in the
\KEPLER\ run, but after all nuclear burning had ceased. The 1D
\KEPLER\ profiles of density, velocity, temperature, and composition
are mapped onto the 2D cylindrical grid of \CASTRO, using the scheme
of \citet{Che13}, which conservatively maps mass, momentum, energy,
isotope compositions from 1D profiles onto multi-dimensional grids.

The \CASTRO\ simulation carried only an octant of the star. The
computational domain was about 10 -- 60 times the radius of the
initial expanding ejecta. As is necessary for Eulerian codes, an
artificial circumstellar medium (CSM) was included in the
calculation. This medium had a density profile $\rho =\rho_0
(r/r_0)^{-3.1}$ where $\rho_0$ is the density at the radius of the
initial expanding ejecta, $r_0$. In the 1 ms model $\rho_0$ and $r_0$
were $2.11 \times 10^{-3} \gcc$ and $1.75\times 10^{11}$ cm,
respectively. In the 5 ms model they were $5.31 \times 10^{-9} \gcc$
and $1.54\times 10^{13}$ cm.  The CSM densities were extended from the
edge of expanding SN ejecta and are much greater than would be
characteristic of any reasonable pre-explosive mass loss rate. They
are more like what might have existed had the core been inside of a
supergiant star. For WR stars with $\dot{M} = 10^{-4}$ $\Msun$
yr$^{-1}$ and escape velocity $1 \times 10^{8}$ cm s$^{-1}$, the CSM
density at $10^{12}$ cm would be $\rho = 5 \times 10^{-12} \gcc$. In
the entire simulation domain, the total mass of this artificial medium
was 0.15 \Msun\ for the 1 ms run and 0.37 \Msun\ for the 5 ms run.
This CSM was used solely to maintain computational stability.  The
density falls off rapidly above the edge of the presupernova star. If
the shock wave generated by magnetar energy deposition passed though a
region where the density fell off more slowly that $r^{-3}$, a reverse
shock would develop. This medium was constructed to decline rapidly
enough in density to avoid this happening.  There are such regions in
the envelopes of both blue and red supergiants
\citep[e.g.][]{Woo95,Her94} and including such envelopes would result
in more mixing and fallback than calculated here \citep{Jog10,Che14a}.

The \CASTRO\ grid, at its coarsest level, had $256\times256$ zones.
Six levels of adaptive mesh refinement were employed for an additional
resolution of up to 64 (2$^6$). This degree of refinement was necessary
to spatially resolve the energy deposition region, as well as the
emergent fluid instabilities. This level of AMR implied an effective
resolution of $16,384\times16,384$. The grid refinement criteria were
based on gradients of density, velocity, and pressure. The hierarchy
nested grids were also constructed in such a way that the energy
deposition region near the magnetar was well resolved.  Reflecting and
outflow boundary conditions were set on the inner and outer boundaries
in both $r$ and $z$, respectively.  A monopole approximation for
self-gravity was included in which a 1D profile of gravitational force
was constructed from the radial average of the density, then the
gravitational field stress of each grid was calculated by the linear
interpolation of the 1D profile. A point-like gravitational source of
1.45 \Msun\ represented the magnetar.

When the 2D simulation began, 100 s after the initial core collapse,
the initial shock from the 1.2 B explosion had already exited the
surface of the star at $r \approx 2\times10^{11}$ cm and the entire
star had nearly reached its terminal velocity (\Fig{presn}). A magnetar
was then introduced with a power described by Eg.\ref{eq:magP} and
Eg.\ref{eq:evo}.  Unlike the \KEPLER\ run which deposited the energy
at a constant rate per unit mass, since \CASTRO\ uses an Eulerian
grid, the energy deposition for the 2D models was a constant {\sl per
  unit volume}. The mass of the 10 zones into which energy is
deposited in the \KEPLER\ run was 0.12 \Msun. In the 2D
\CASTRO\ simulations, the power is deposited in a constant volume
bounded by a radius of about corresponding to $2\%$ of the initial
radius of ejecta.. In no case was the instantaneous kinetic energy in
this material an appreciable fraction of the supernova explosion
energy, i.e., most of the energy deposited went into doing work where
the wind terminated.

%%%%%%%%%%%%%%%%%%%%%%%%%%%%%%%%%%%%%%%%%%%%%%%%
% Section 3 - Results %
%%%%%%%%%%%%%%%%%%%%%%%%%%%%%%%%%%%%%%%%%%%%%%%%
\section{RESULTS}
\lSect{results}

Two sets of calculations were carried out, each using both
\KEPLER\ and \CASTRO\ to model a given explosion in 1D and 2D. In both
cases, a constant magnetar field strength of $4 \times 10^{14}$ G was
assumed, but the studies used initial rotational periods of 1 ms and 5
ms, corresponding to initial rotational energies of $\rm20$ B and
$\rm0.8$ B respectively. Ideally, we would have liked to run both
simulations for at least one magnetar spin-down time scale, $\rm t_m$,
which is 12,500 s and 312,000 s for the 1 ms and 5 ms cases
respectively. Both \KEPLER\ calculations were run to about 300 days
and satisfied this condition. In 2D however, owing to the small
timesteps in the finest zones ($\sim 10^{-3}$ s for the 1 ms run and
$\sim 10^{-1}$ s for the 5 ms run), we were only able to simulate the
first 3,000 s (50 min) of the 1 ms case and the first 520,000 s (6
days) of the 5 ms case. During these times $\rm4.8$ B out of the
available $\rm20$ B is deposited in the high energy run and $\rm0.5$ B
of the available $\rm0.8$ B deposited for the low energy run. Energy
deposition in the high energy case was therefore far from over when
the calculation was stopped, while the low energy case was essentially
complete. These two cases were selected to represent situations where
the deposited energy greatly exceeded or was substantially less than
the initial dialed in supernova energy, $\rm 1.2B$ and the qualitative
results will not be altered by this inadequacy. One should keep in
mind though that the high energy model would have mixed even more than
calculated here.  A recently discovered transient, ASASSN-15lh is
possibly explained by a magnetar of rotational energy $\approx 40$ B
\citep{Met15,Ber16,Suk16}, which resembles the 1 ms run here.

\subsection{One-Dimensional Results}
\lSect{kepler_R}

The bolometric light curves calculated using \KEPLER\ are shown in
\Fig{lc}. The 1 ms model produces a light curve that agrees well with
the observations of the transient PTF10cwr \citep{Qui11}, while the 5 ms
model roughly fits the measurements for the transient PTF11rks
\citep{Ins13}. We do not include the ASASSN-15lh light curve here.  
	Because \citet{Suk16} have given a light curve calculated for ASASSN-15lh by using \KEPLER, 
	and we are also waiting to see if the identification of this object as a supernova persists.

A density spike emerges in both calculations beginning $\approx
100$ sec after the magnetar is turned on in the 1 ms model run and
after $\approx 10,000$ sec in the 5 ms run.  The amplitude of this
spike grows with time and eventually includes most of the ejected
mass. The density spike in the 1 ms model, shown in \Fig{rho1}, grows
to have a contrast of over three orders of magnitude with surrounding
ejecta during the first 1,000 seconds. The density spike in the 5 ms
model has a similar evolution, but grows on a longer time scale of
hours. In past multi-dimensional simulations, similar spikes have been
the location of fluid instabilities \citep[e.g.][]{Che92,Jun98,Blo01}.

\subsection{Two-Dimensional Results}
\lSect{castro_R}

The \CASTRO\ calculations were begun from the \KEPLER\ 1D model at
$\approx 100$ sec for the 1 ms run and $\approx 10,000$ sec for the 5
ms run.  By this time, the original expanding ejecta had
reached a radius of $1.75 \times 10^{11}$ cm and $ \sim 1.54 \times
10^{13}$ cm, respectively.  The simulated domain used $r = 2.5
\times10^{12}$ cm for the 1 ms model, and $r = 1\times10^{15}$ cm for
the 5 ms model, with the finest zones being about $1.2\times10^8$ cm,
and $6.1\times10^{10}$ cm. The size of the domain determined the
duration of the simulation.  The difference in assumed magnetar
rotation rates results in different energy deposition rates that cause
the emerging spike to appear at very different times.  For the 1 ms
model, a much faster and more vigorous interaction with the overlying
ejecta was observed. The 5 ms model had to be evolved much longer
since the energy was deposited over a longer time. Although these were
only 2D simulations, small time steps required by the fine spatial
resolution still made they very computationally expensive.  The 1 ms
model took about 360,000 CPU hours on $\it Hopper$ and the 5 ms model
took 280,000 CPU hours on $\it Edison$ at the National Energy Research
Scientific Computing Center (NERSC).

% Fig 3
\begin{figure}[h]
\centering
\includegraphics[width=\columnwidth]{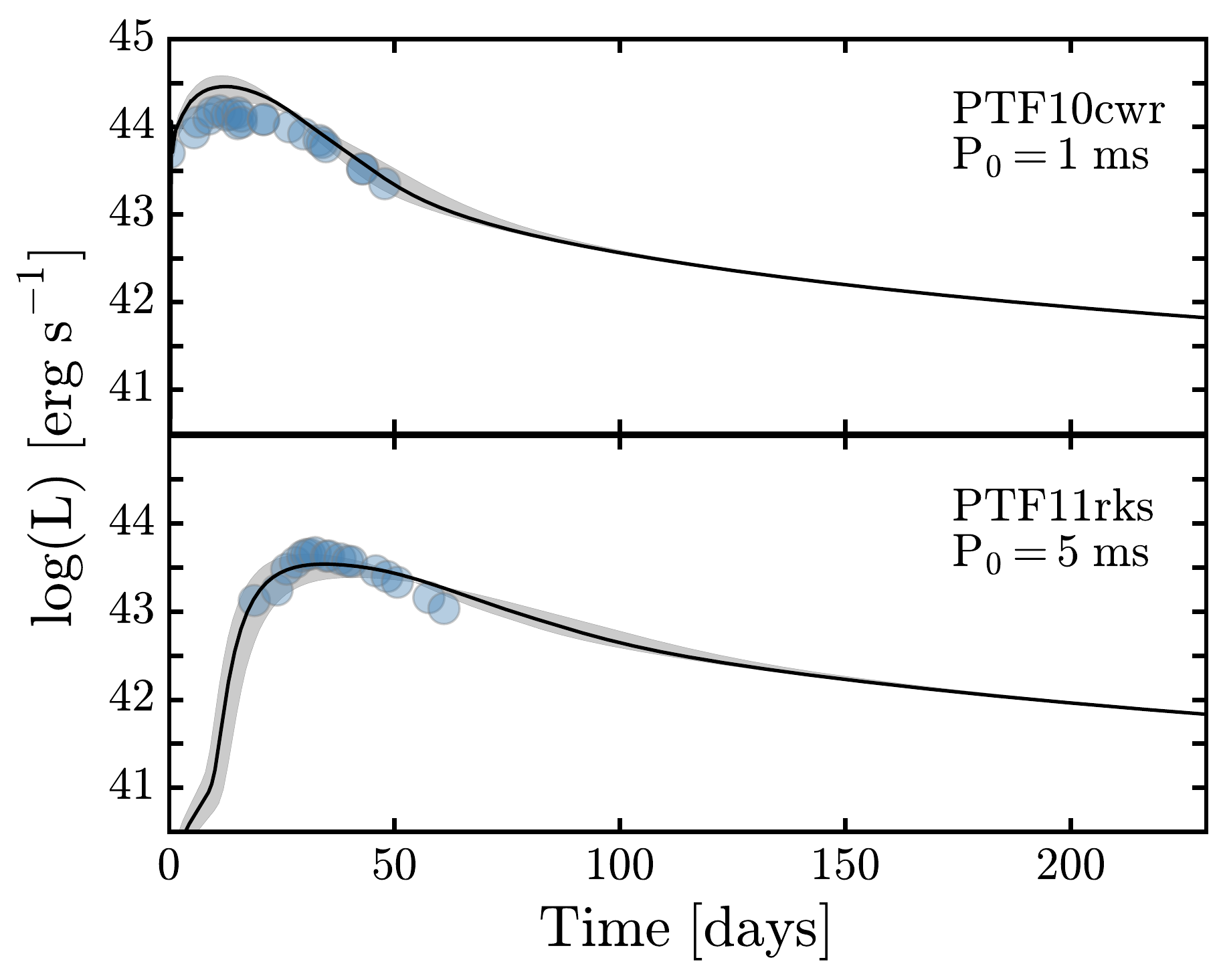} 
\caption{Bolometric light curves of PTF10cwr and PTF11rks (blue
  circles), compared with 1D model light curves calculated for the
  explosion shown in \Fig{presn}, but including an embedded magnetar
  with an initial rotational period of 1 ms or 5 ms, and a magnetic
  field strength of $4 \times 10^{14}$ G. Both light curves were
  calculated in 1D using the \KEPLER\ code and the results are sensitive
  to the assumed opacity. Gray ranges in both panels indicate choices
  of opacity $\rm\kappa=0.05-0.2\ g cm^{-2}$. The black curves assumed
  $\rm\kappa=0.1\ g\ cm^{-2}$.  \lFig{lc}}
\end{figure}

% Fig 4
\begin{figure}[h]
\centering
\includegraphics[width=\columnwidth]{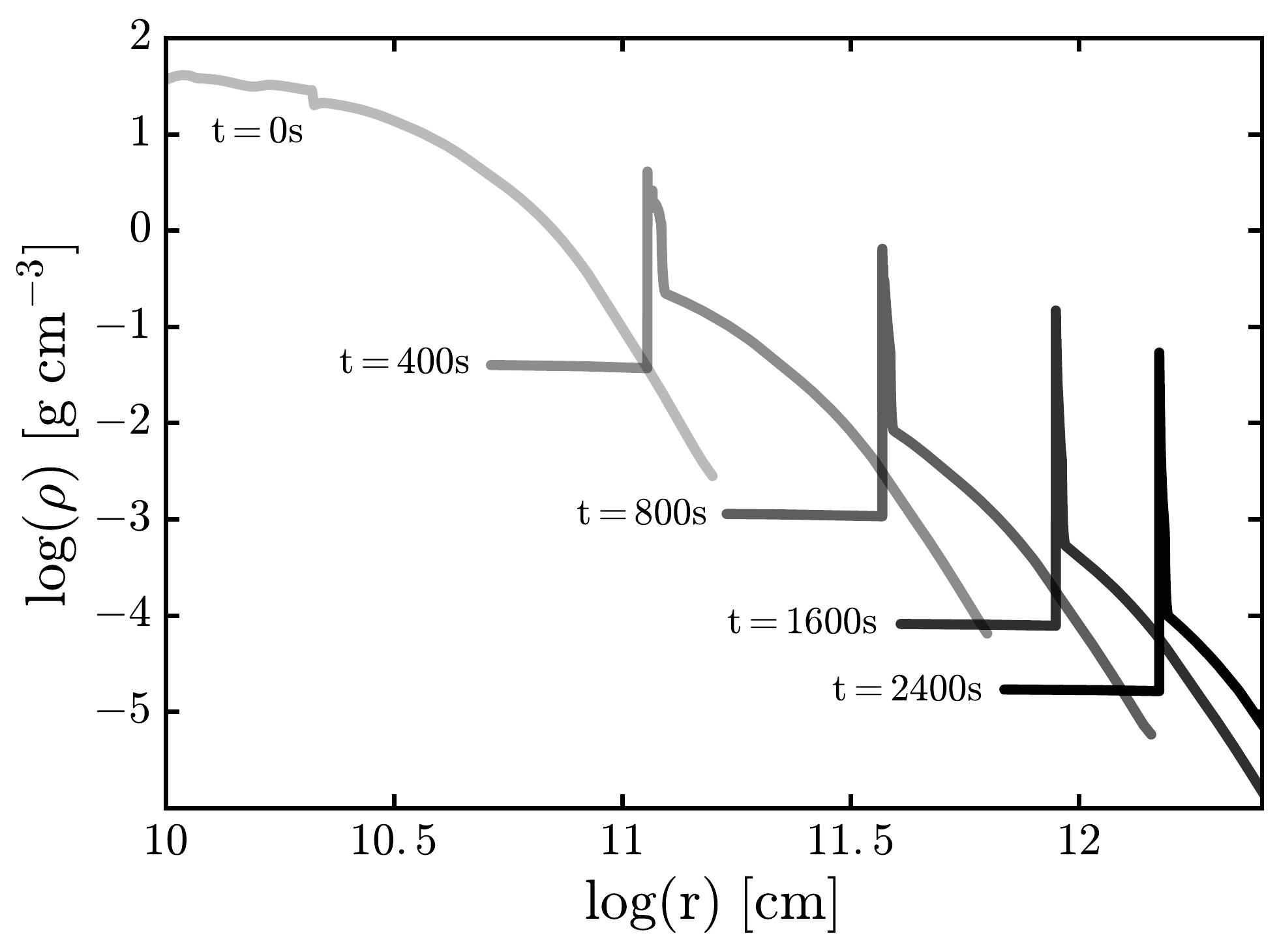} 
\caption{Early evolution of the density profiles from the 1 ms
  magnetar as calculated in 1D using the \KEPLER\ code. Times are
  measured since the magnetar deposition was turned on, i.e., time
  since explosion minus 100 s. A prominent density spike starts to
  emerge at $\approx100$ s as the wind of the magnetar snowplows into
  overlying ejecta. The amplitude of the spike grows rapidly with time
  and eventually includes most of the ejected mass. \lFig{rho1}}
\end{figure}

% Fig 5
\begin{figure}[h]
\centering
\includegraphics[width=\columnwidth]{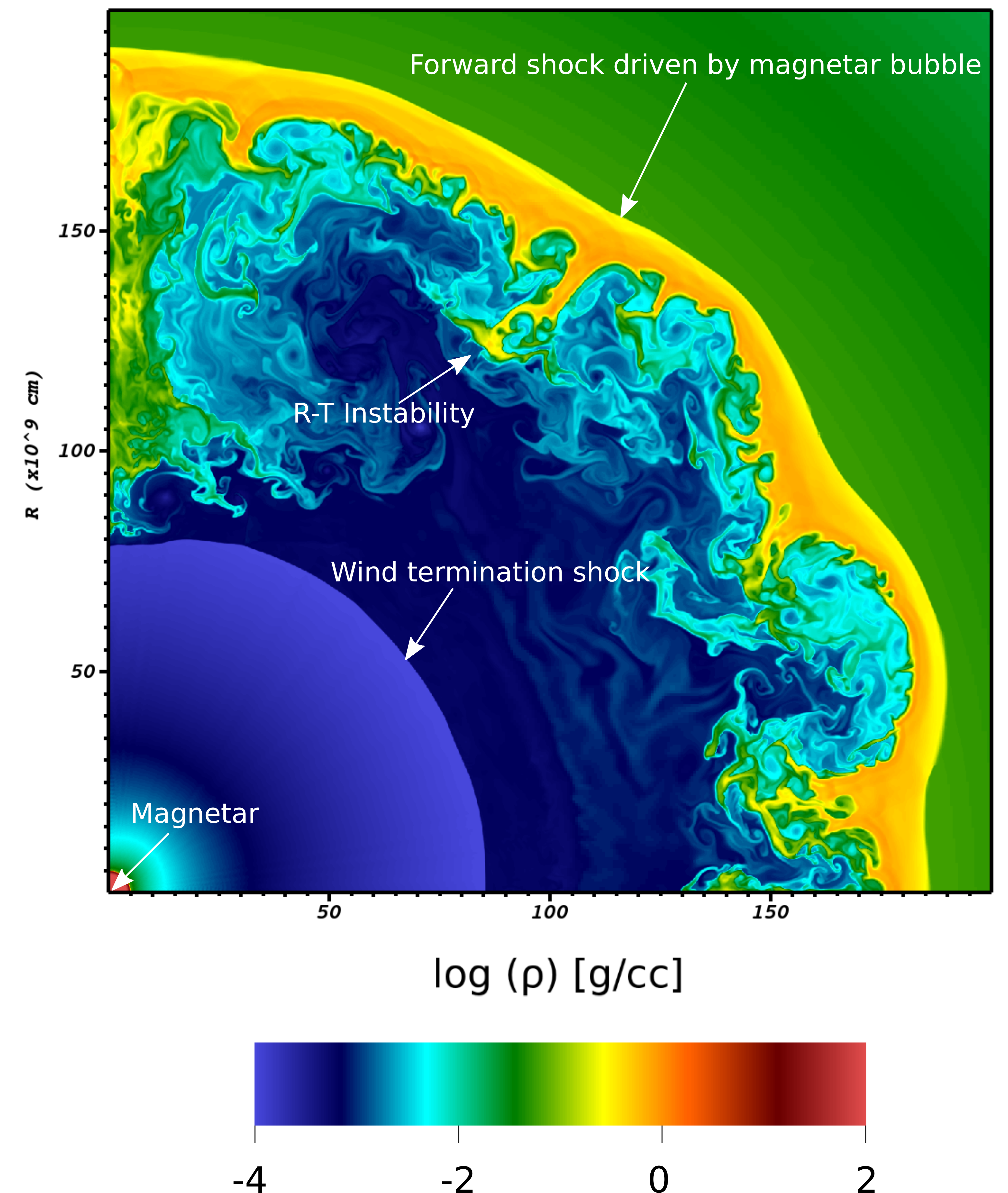} 
\caption{Density structure in the inner $2 \times 10^{11}$ cm of the
  supernova for the 1 ms model at an early time in the 2D
  calculation. The time is 600 s after the magnetar was turned on and
  the boundary of the mixed region, also called the ``radiation
  bubble'', extends to about $1.8 \times 10^{11}$ cm .  The boundary
  of smooth blue region inside $8.5 \times 10^{10}$ cm marks the
  termination of the supersonic magnetar wind as it slams into the
  slower moving overlying ejecta. Later (e.g., \Fig{vel}), this
  termination shock becomes more irregular as the wind begins to break
  through. The ram pressure of the rapidly moving wind accelerates the
  overlying matter causing a pile up of density which increases
  roughly monotonically with radius between the wind termination shock
  and the forward shock. Accelerating this density inversion causes a
  Rayleigh-Taylor instability that mixes the region between the two
  shocks. This mixed region is not present in the 1D simulation
  (\Fig{rho1}) where there is only one shock. Farther out, this
  mixture of radiation, wind, and supernova ejecta plows into slower
  moving supernova ejecta material resulting in a dense ``pile-up''
  bounded by a second shock. The boundary of original supernova surface
  is at $\sim 1.1 \times 10^{12}$ cm, well outside the region plotted.
  \lFig{break}}
\end{figure}

\subsubsection{Formation of a 2D radiative bubble}
\lSect{bubble}

The energy injected by the magnetar heats the surrounding gas, causing
it to expand and reach high speed. If the mass of the energy
deposition region in \KEPLER\ or the size of the energy deposition
region in \CASTRO\ had been too small, super-luminal motion would
have resulted. Mass was therefore added, along with the energy in the
\CASTRO\ calculation so as to allow a high velocity wind, but prevent
expansion faster than the speed of light. Had more
mass been added, it would have moved with slower speed, but the work
done at the wind termination shock, $\rho v^2$ times the change in
volume, would have been the same.  The mass addition rates employed in
\CASTRO\ were $2.5\times 10^{-6} \Msun\ s^{-1}$ for the 1 ms run and
$1.2\times 10^{-9} \Msun\ s^{-1}$ for the 5 ms run. During the entire
run the accumulated mass was about $7.5\times10^{-3} \Msun$ for the 1
ms run and $6.2 \times10^{-4} \Msun$ for the 5 ms run. Both values are
negligible compared with the mass of the supernova ejecta.

In both the 1D and 2D calculations the gas energized by the magnetar
pushes the overlying cooler material ahead of it and forms a dense
shell. In 1D, there is no possibility for mixing so the shell is
stable. The ``termination shock'' at the edge of the 10 Lagrangian
shells where energy is deposited is located at almost the same radius
as the ``forward shock'', essentially the leading edge of the density
pile up. There is effectively only one shock and all the swept up
matter is compressed within it.

In 2D, however, the region inside the maximum density is unstable and
mixes. A growing region develops between the wind termination shock
and the forward shock where the acceleration caused by the wind
operates in a region of decreasing density (\Fig{break}). Near the
forward shock and beyond there is no density inversion and no mixing,
but behind it the ejecta is Rayleigh-Taylor unstable. The two shocks
separate and between them the star is mixed and a convoluted density
structure develops. We shall refer to the entire region of mixed
matter and wind behind the forward shock as the ``bubble''. For the 1
ms model, the bubble expands at the rate of $2-5\times10^{8}$ cm
s$^{-1}$, taking about 600 s to grow to the size of the original
progenitor star, $\approx 2\times 10^{11}$ cm (\Fig{break}).  Later,
for the 1 ms case, the bubble expands even faster, $> 10^{9}$ cm
s$^{-1}$, and its leading edge starts to catch up with the outer edge
of the original supernova.

Even though the ejected matter becomes mixed in 2D, it is still
largely concentrated in the outer part of the bubble.  The supernova
is ``hollow'' and shellular with a thickness much less than its
radius.  So long as the energy being dumped in by the magnetar is
comparable to the kinetic energy of the matter outside the bubble, the
mixing and compression continue. 

''Bubble breakout'' is defined as the point when an appreciable part
of the magnetar-accelerated shell first reaches a speed comparable to
that of the fastest expanding ejecta.  From \Fig{presn}, only
$2.7\times10^{-3}$\Msun\ moves faster than $2 \times 10^{9}$ cm
s$^{-1}$ in the original SN, so mixed material that attains that speed
has clearly escaped. This is defined as the condition of {\sl ``strong
  breakout''}. A larger, but still small amount of ejecta, 0.17 \Msun,
moves faster than $1 \times 10^{9}$ cm s$^{-1}$, and matching this
speed defines {\sl ``weak breakout''}.  More specifically, weak
breakout occurs when a magnetar-accelerated shell mass of 0.17
\Msun\ moves faster than $1 \times 10^{9}$ cm s$^{-1}$.  Once this
breakout occurs, the dense shell become fragmented and opens gaps that
allow hot trapped magnetar wind, and eventually the magnetar radiation
itself to escape.  For the 1 ms case, this condition implied the full
mixing of the entire explosion. The thick shell of bubble breaks
though the original surface of the star as shown in \Fig{2d1}(c) (d).
About $1.89\Msun$ of the bubble has reached the weak breakout phase
2,400 sec after the explosion and more would continue had the
calculation been run longer. The break out of the bubble in the 5 ms
model (\Fig{2d2}) is less extreme.  In this case, fragmentation
and mixing is not as developed, but about 0.24 \Msun\ of the bubble
barely reaches weak breakout with a $25 -30 ^{\circ}$ opening angle
roughly 5 days after the explosion.

The dynamics of the breakout thus depends upon the relative energy
input by the central magnetar and the original explosion.  Breakout
can happen, in principle, when the amount of energy deposited by the
magnetar, $\int \rm{L_m} dt$, and expended in doing PdV work at the
termination shock, becomes comparable with the original kinetic energy
of the ejecta. This is an approximate condition though. Neither the
original supernova nor the magnetar-accelerated bubble move at uniform
speed. The original supernova has very high speed at its edge due to
shock steepening at breakout (\Fig{presn}).  The bubble has variable
speeds at different angles and, at late times, is honeycombed by
magnetar wind that has broken though the termination shock
(\Fig{vel}).

Nevertheless, the calculations suggest a breakout time, $t_b$, of
roughly $\rm{E_{sn}/L_{m}}$, where $\rm{E_{sn}}$ is the kinetic energy
of the original supernova and $\rm{L_{m}}$ is the magnetar luminosity.
For the 1 ms case, this gives $1.2\times10^{51}/10^{48} \sim 1200$ s
which is consistent with the results of the simulation.  For the 5 ms
run. the total energy deposited by the magnetar is 0.8 B which close
to $\rm{E_{sn}}$. Only a fraction of the shell experiences
breakout. Once the shell starts to fragment, however, the piston doing
the PdV work is less efficient. Gaps are opened for the hot gas to
break out.  For the 1 ms model at the post breakout phase ＄$t \sim
2400$ sec, $69\%$ of magnetar energy went to radiation and $31\%$ to
accelerating the ejecta.  It is possible that the dipole radiation of
the magnetar may be able to escape through the holes formed by the
bubble as it breaks out \citep[see also][]{Met14,Met15}, especially
since the calculation followed only a fraction of the energy
deposition. Such radiation breakout may be a common occurrence in
energetic magnetar-powered supernovae suggested \citet{Kas10}. More calculations
including radiation transport and a realistic spectrum for the
magnetar should be done in two dimensions to better determine the
observable consequences of breakout.

% Fig 6
\begin{figure*}[h]
\centering
\includegraphics[width=\textwidth]{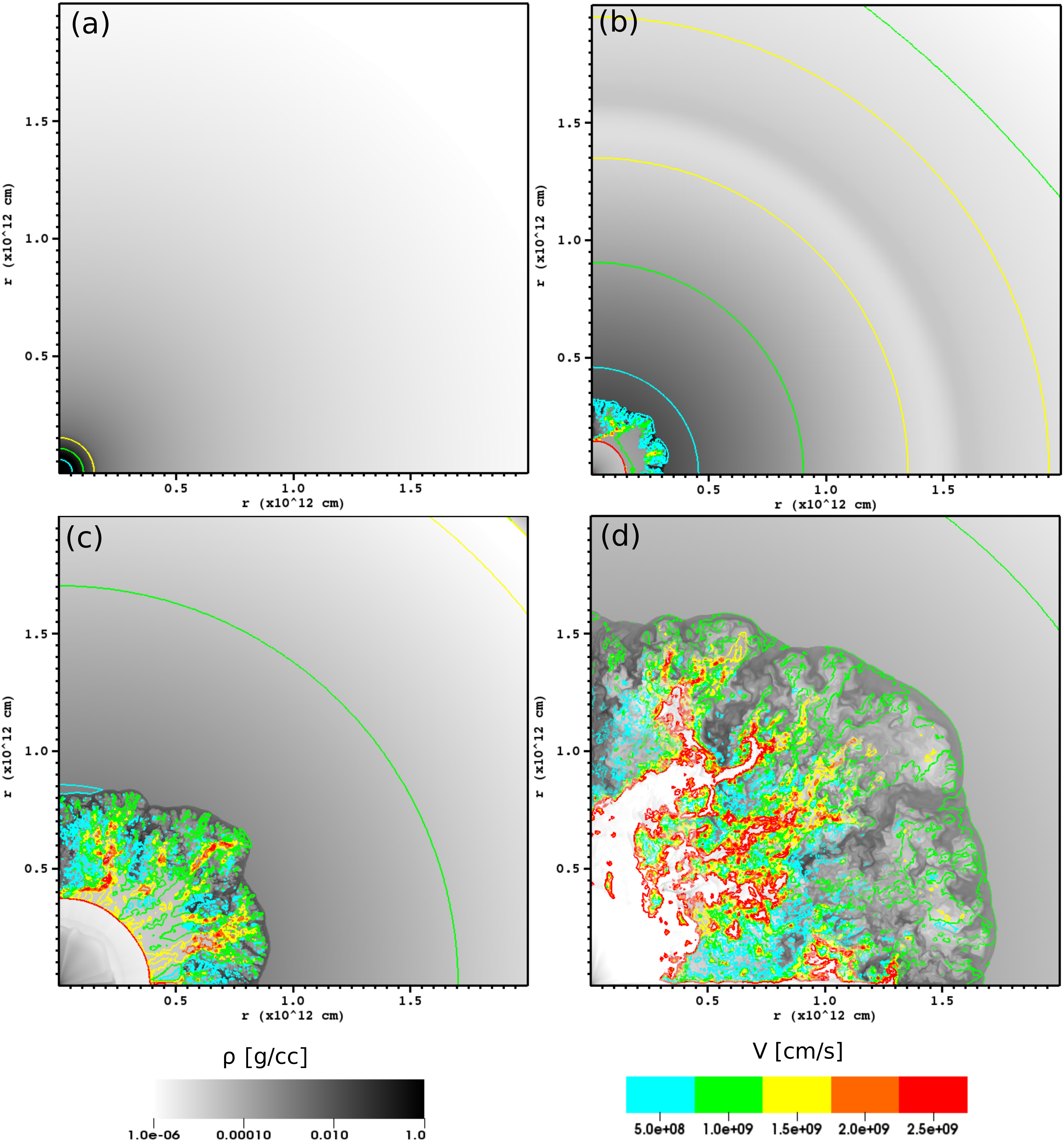} 
\caption{Evolution of the mixed region in the early phases of the 1 ms
  model. Color coding and contours show the densities and velocities
  of SN ejecta.  Panels (a) - (d) are at 0, 800, 1,600, and 2,400
  seconds, respectively.  After the magnetar begins to deposit energy,
  fluid instabilities develop from tiny fingers as shown in Panel (b)
  and a thin shell has formed. The shell is promptly accelerated by
  the high-speed magnetar wind.  In Panel (c), some fraction of shell
  and the gas behind it has exceeded $1 \times 10^{9}$ cm s$^{-1}$
  . In Panel (d), the entire shell has exceed $1 \times 10^{9}$ cm
  s$^{-1}$ and the week breakout has occurred.  It is expected that
  the strong breakout ( v $\ge 2 \times 10^{9}$ cm s$^{-1}$) will
  occur shortly. Velocities in the low density region (white area) in Panel (d) have exceeded 
  10$^{10}$ cm s$^{-1}$
  \lFig{2d1}}
\end{figure*}

% Fig 7
\begin{figure*}[h]
\centering
\includegraphics[width=\textwidth]{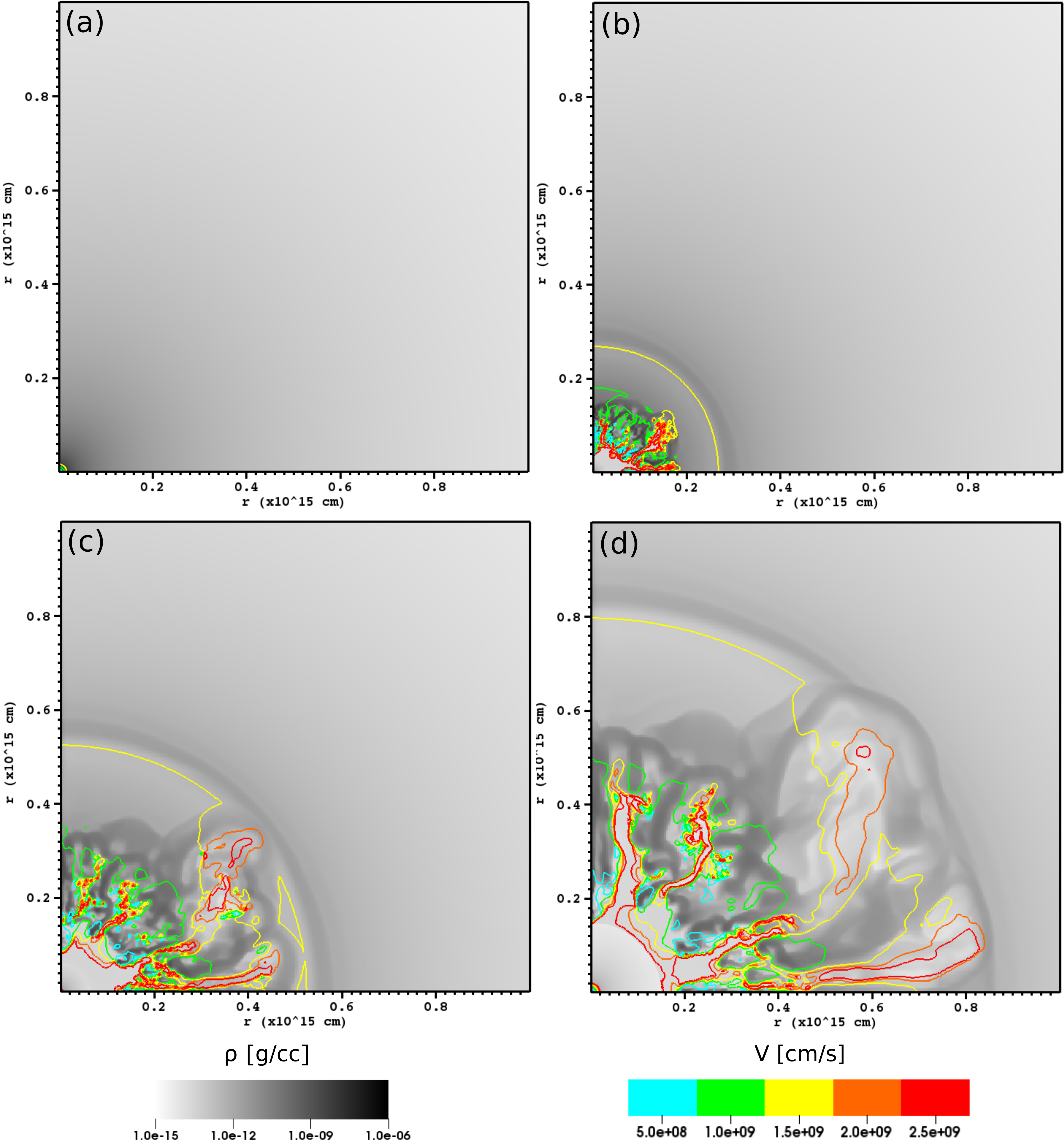}
\caption{Similar to \Fig{2d1} but for the 5 ms model.  Panels (a) -
  (d) now show the densities and velocities at 0, 2, 4, and 6 days,
  respectively.  In Panel (b), fluid instabilities again appear and
  two prominent fingers have formed. The dense shell is eventually
  penetrated by these two fingers (Panel (d)).  Only the dilute gas
  behind the shells exceeds $1 \times 10^{9}$ cm s$^{-1}$. The shell
  itself does not move faster because of the smaller energy deposited
  and most of the shell will not break out.  The overall mixing and
  fragmentation is less extensive in comparison with the 1 ms
  model. \lFig{2d2}}
\end{figure*}

% Fig 8
\begin{figure}[h]
\centering
\includegraphics[width=\columnwidth]{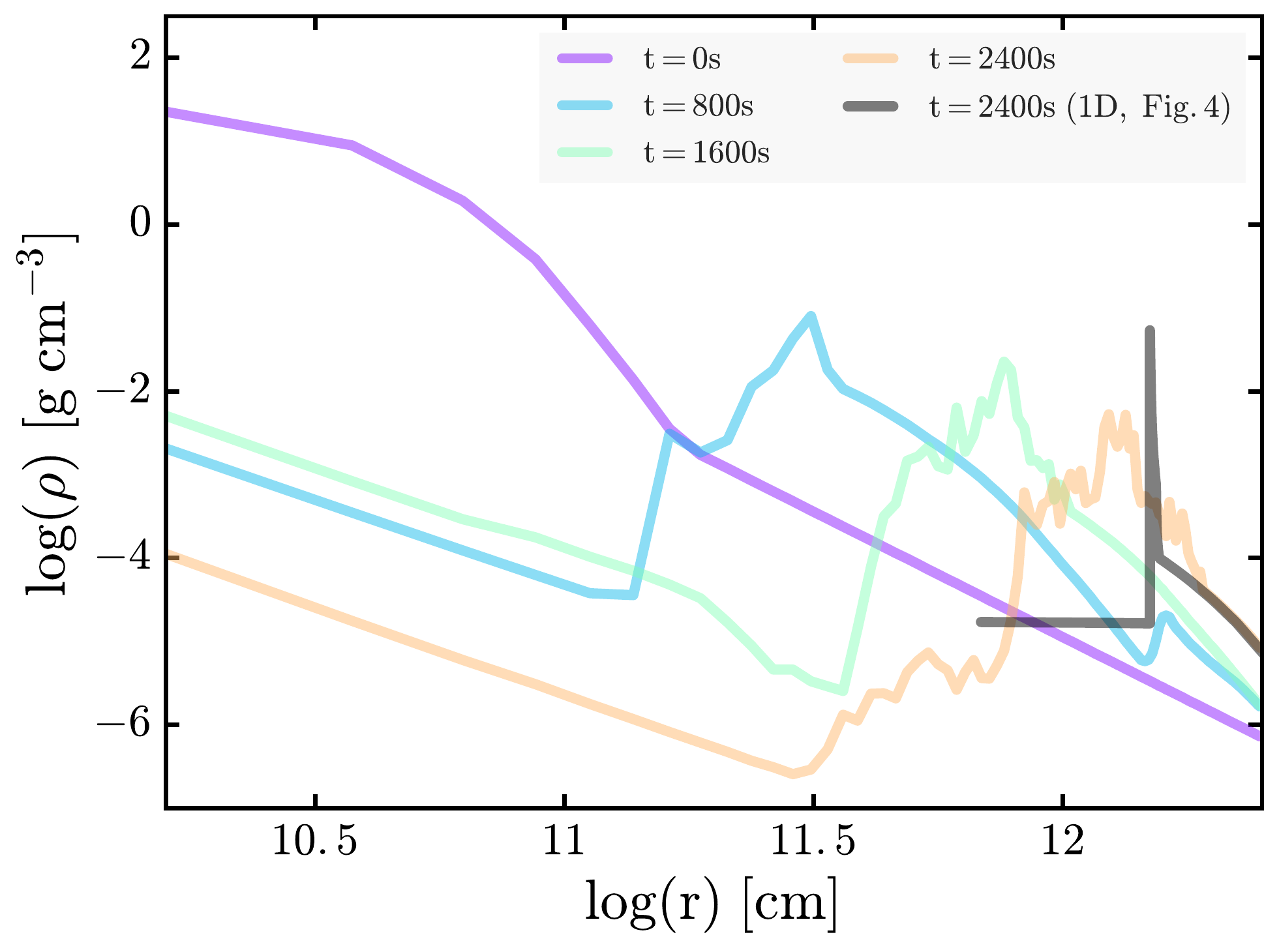} 
\caption{Angle-averaged density profiles of 1 ms case, as calculated
  in the 2D \CASTRO\ model. Curves represent profile snapshots shown
  in \Fig{2d1}. The fluid instabilities in 2D smear out the density
  spike seen in 1D and cause mixing. The profile from the 1D
  \KEPLER\ run (black curve, \Fig{rho1}) at 2400 s is also shown for
  comparison. Note that even in 2D the supernova is still
  ``shellular'' with a hollow center.  \lFig{rho2}}
\end{figure}

\subsubsection{Evolution to the Coasting Phase}
\lSect{coast}

Once the bubble breaks out of the expanding supernova ejecta, it runs
into the artificial circumstellar medium (CSM).  The deformed
structure continues evolving as the bubble expands,
however. Significant mixing has occurred inside the bubble and broken
its spherical symmetry.

For the 1 ms model shown in (d) of \Fig{2d1}, weak breakout has
occured and strong breakout will follow shortly. After this time,
the evolution of fluid instabilities slows down, but will still
continue since only about $20\%$ of magnetar energy has been
deposited. An increasing fraction of the injected energy would
presumably escape from the perforated shell with a spectrum that might
eventually resemble that of the pulsar \citep{Kas10} .

The angle-averaged profiles of density are shown in \Fig{rho2}.
Spikes seen in the 1D \KEPLER\ models do not disappear in 2D, but are
substantially eroded and broadened. Fluid instabilities in the 2D
study result in a ``noisy bump'' when angle averaged, but in fact the
shell is being broken up and mixed. The relative density constraint,
$\delta \rho = {\rho-\langle\rho\rangle/\rho}$ is 10 to 100
within the mixing region rather than up to $10^3$ as seen in the 1D
study. In 1D models, most radiation is emitted from the density spike,
which suggests the radiation may continue to come from the mixed
region in the multidimensional models since that is where most of the
matter is.

Due to the continuing injection of energy by the central magnetar, the
structure continues to evolve. When the bubble expands to a large
radius ($>$ ten times radius of initial expanding ejecta), the ejecta
are still not expanding fully homologously since the internal energy
of gas still exceeds $10\%$ of its kinetic energy, but a filamentary
structure of the ejecta has been determined (\Fig{2db}). The 1D
angle-averaged abundances are shown in \Fig{spec} for the 1 ms model.
The major mixing occurred at region of fragmented dense shell and some
fraction of \Ni\ appears at the outer edge of the fragmented shell. If
such dredging up of \Ni\ indeed happens at an early phase of magnetar
evolution, there is the possibility of early gamma-ray detection from
the \Ni\ decay in the local or nearby galaxies. The \Ni\ would leave
footprints on the magnetar-powered SN remnant and might resemble the
iron observed on the outskirts of the Cas A SN remnant \citep{Vin08}.

% Fig 9
\begin{figure}[h]
	\centering
	\includegraphics[width=\columnwidth]{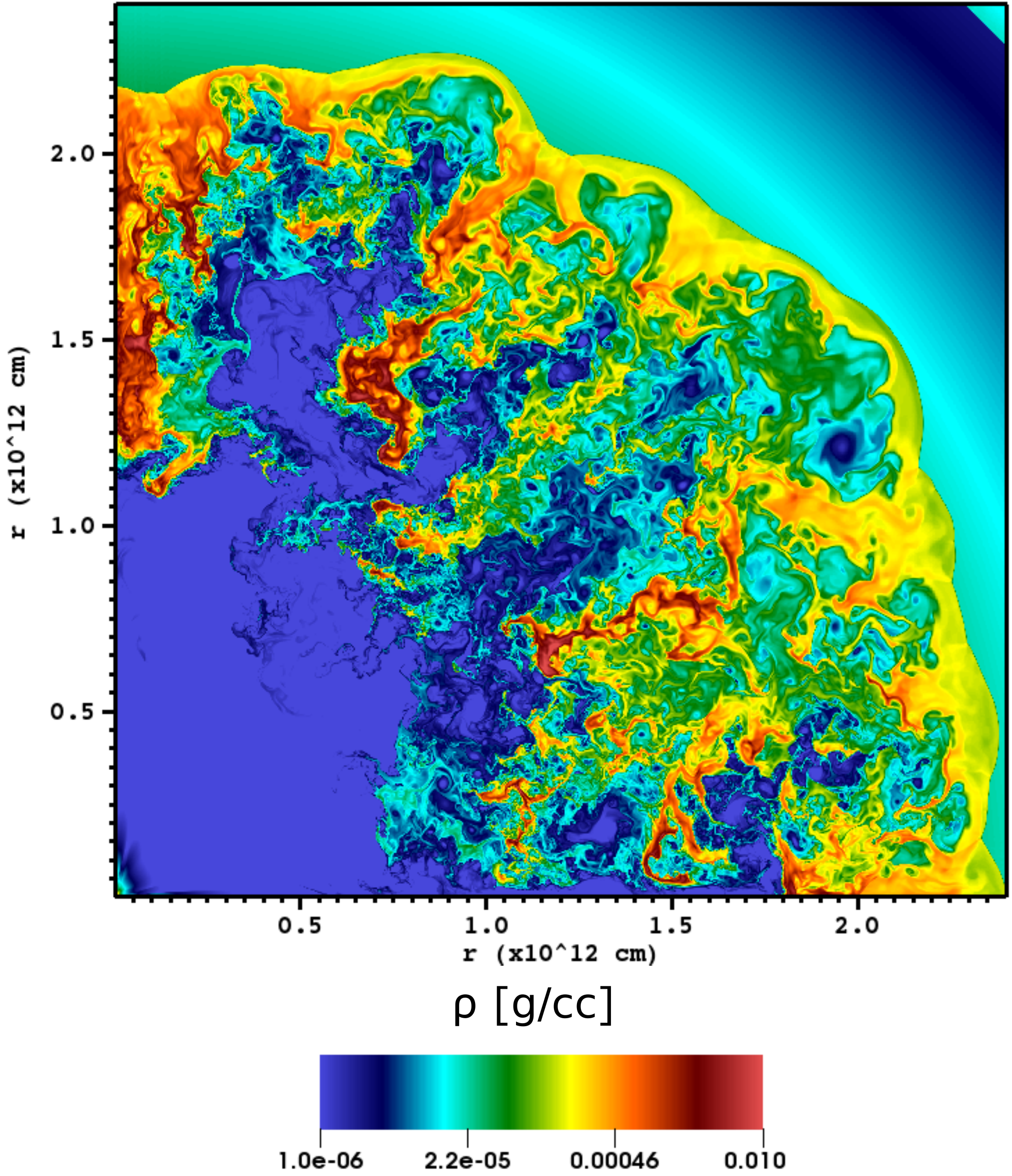} 
	\caption{Mixing of the 1 ms model is shown for the last model
          calculated, $t \approx 3,000$ s. Density is given on a
          logarithmic scale from 10$^{-6}$ g cm$^{-3}$ to
          10$^{-2}$ g cm$^{-3}$. Regions of low density are
          also regions of high expansion speed (\Fig{vel}). The highly
          fractured nature of the mixed ejecta will alter its
          observational signatures and the structure of the supernova
          remnant. \lFig{2db}}
\end{figure}

% Fig 10
\begin{figure}[h]
\centering
\includegraphics[width=\columnwidth]{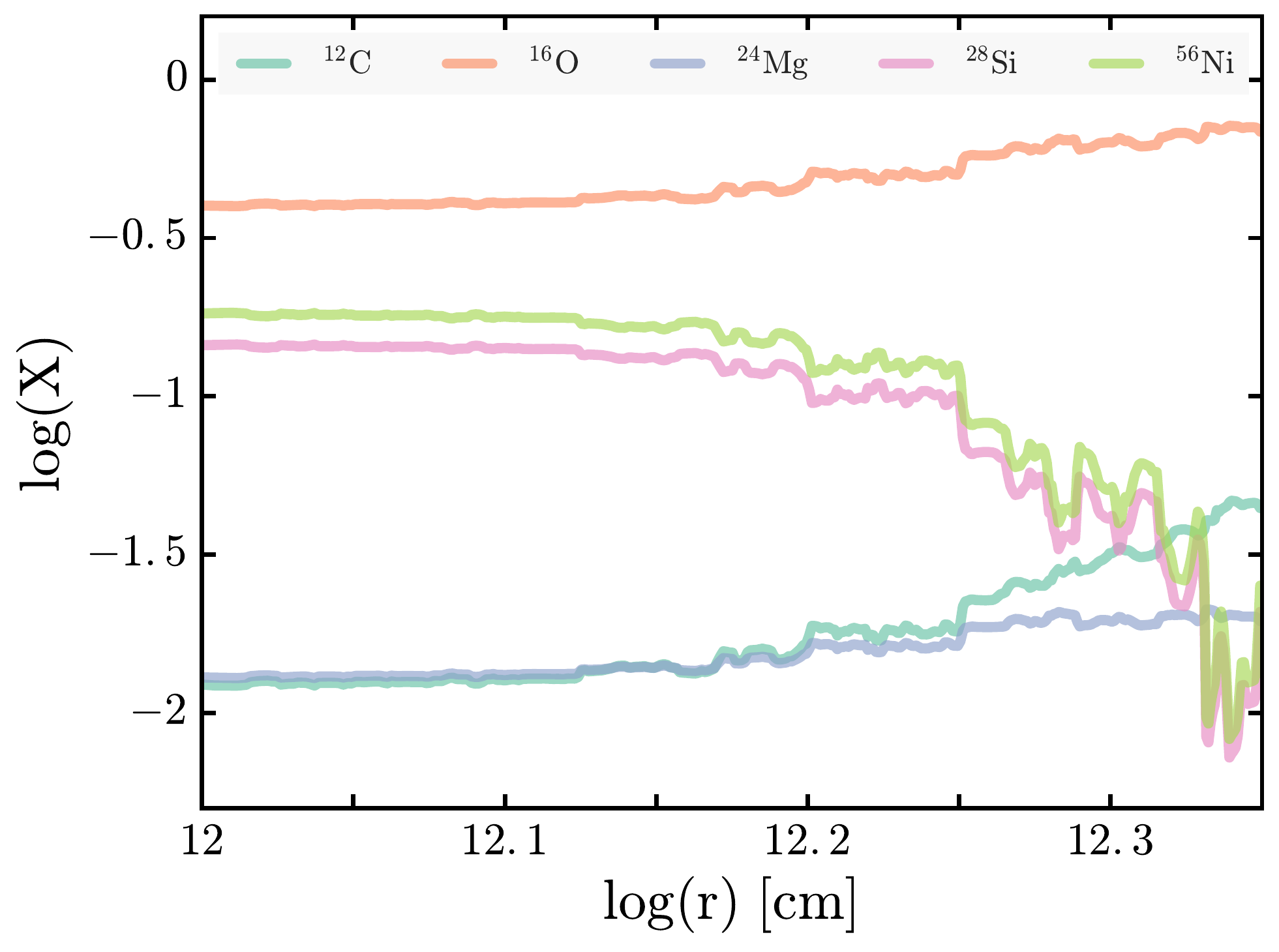} 
\caption{Angle-averaged elemental abundances for the 1 ms model at $t
  = 3,000$ s.  Some \Ni\ from the interior of star has been mixed out
  and enriches the outskirts of the bubble with a mass fraction of
  about 0.01. This will leave a compositional imprint on the
  composition distribution in the supernova remnant.  \lFig{spec}}
\end{figure}

% Fig 11
\begin{figure}[h]
	\centering
	\includegraphics[width=\columnwidth]{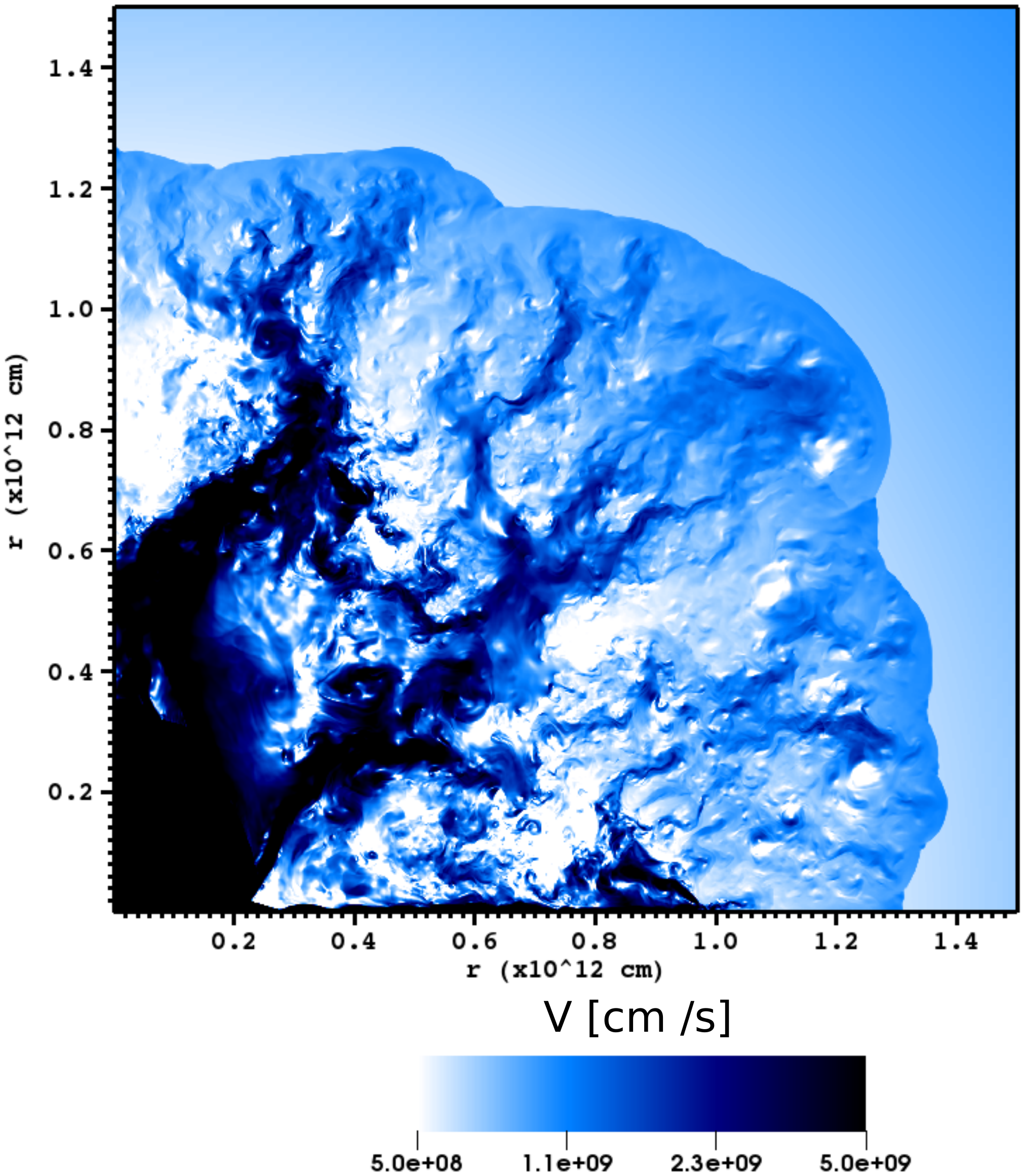} 
	\caption{Ejecta velocity in the 1 ms model at $t \approx
          2,000$ s, near weak breakout.  The color scale indicates a
          logarithmic scale from $5\times10^8$ cm s$^{-1}$ to $5
          \times 10^{9}$ cm s$^{-1}$. There is a clear interface
          between the bubble and its surroundings that marks the
          location of a shock front. Within the bubble, expansion
          speeds approach a fraction of light speed. Farther out,
          channels of high speed ejecta are opening up.  At the time
          shown only $\rm3.2$ B of the available $\rm20$ B has
          deposited and the bubble will achieve strong breakout
          shortly. \lFig{vel}}
\end{figure}

% Fig 12
\begin{figure*}
\begin{center}
\begin{tabular}{cc}
\includegraphics[width=\textwidth]{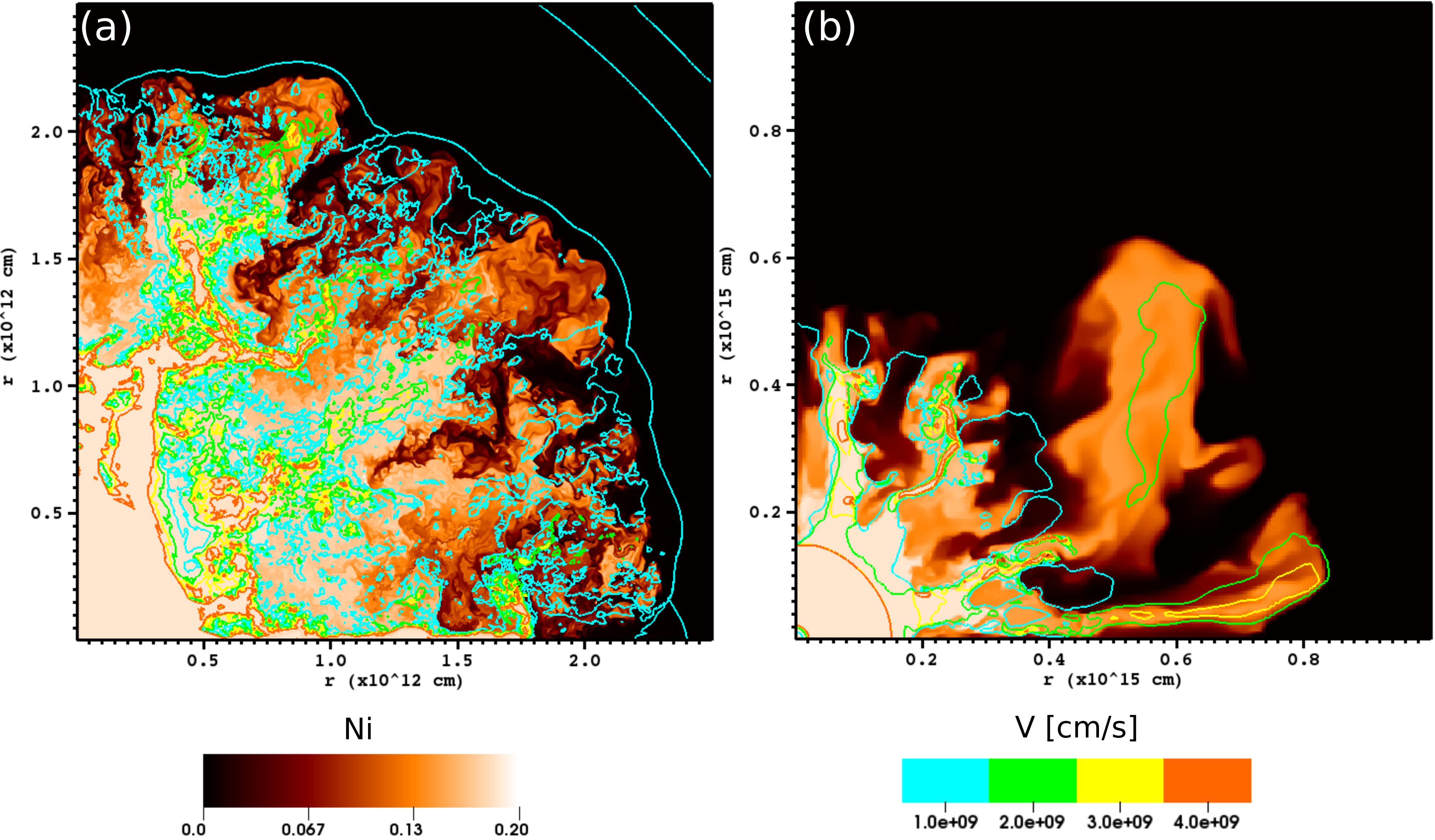}
                        \\
		\end{tabular}
\end{center}
\caption{2D \Ni\ and velocity distributions for the 1 ms model (Panel
  a) and 5 ms model (Panel b) at the last model calculated.  \Ni\ is
  mixed out with velocities of $1 \times 10^{9}$ cm s$^{-1}$. In Panel
  a, the entire shell has reached the weak breakout regime and will
  become optically thin region when the strong breakout occurs.  In
  Panel b, two prominent \Ni\ fingers appear at the breakout phase.
  The mixing of \Ni\ and other chemical elements may be reflected in
  the spectrum.}
\vspace{0.2in}
\lFig{ni}
\end{figure*}

\subsection{Discussion}
\lSect{disc}

\subsubsection{Model Results}

The fluid instabilities of a magnetar-powered supernova are similar to
those previously found for pulsar-wind nebulae
\citep{Che92,Jun98,Blo01}.  Two kinds of instabilities are seen. When
the magnetar first heats the gas and drives an outflow (the ``magnetar
wind'') an instability develops near the contact discontinuity
where dense ejecta is being accelerated. The low density hot gas
colliding with the dense ejecta is RT unstable \Fig{break}.  A number
of long fingers are generated by this instability, and these fingers
are Kelvin-Helmholtz unstable at their boundaries due to their
relative motion with respect to the background flow.

As the shell approaches the boundary of the expanding ejecta, its
expansion rate can be estimated using dimensional analysis
\citep{Jun98}, $r \propto t^{(6-a-b)/(5-a)}$, where $a$ and $b$ are
the power-law indices for the moving ejecta density ($\rho \propto
r^{-a} $) and the magnetar luminosity (${\rm L_m} \propto t^{-b} $)
respectively.  Using the CSM density ($a = 3.1 $) and assuming a
constant magnetar luminosity ($b = 0$), the bubble radius expands
roughly as $r \propto t^{1.52}$. The shell accelerates and expands
supersonically as shown in \Fig{vel}. The second Rayleigh-Taylor
instability is driven by the acceleration of bubble's shell. This is
the nonlinear thin shell instability (NTSI) found by
\citet{Vis94}). It happens when a thin slab bounded by a shock on one
side and a contact discontinuity to a higher temperature region on the
other is subject to a nonlinear instability in which the
perturbation's wavelength is larger than the width of shell.  In our
case, the shell is bounded by the forward shock and relativistic
magnetar wind. The NTSI provides a major mechanism to drive the mixing
and fragmentation formation.
	The original motivation to setup a = 3.1 is to prevent the reverse shock 
	formation so it would not induce additional mixing. In this study, a luminosity source is provided from the central magnetar. The forward shock of expanding bubble is no longer adiabatic.  This shock indeed accelerates both in the constant density CSM (a = 0) and in wind-like ISM (a = 2). Since the growth of rate of NTSI is marginally proportional to the shell velocities \citep{Blo96}.  If 
	we employ the constant CSM or wind-like ISM in our simulations, the overall fragmentation structure  may not  be as evolved as the results we present here.

In the Type I supernova model studied, radiation from sufficiently
energetic magnetars breaks out of the dense layer bounding the
radiative bubble during the NTSI phase and becomes
observable. Depending upon the magnetar spectrum, this emission might
take the form of hard x-rays.  In addition, the mixing driven by the
fluid instabilities alters the dynamics and chemical compositions of
supernovae ejecta. Since mixing is strongest in the region of the flow
from which most of the radiation originates, it will certainly affect
the supernova light curve and spectrum.  As shown in \Fig{ni}, it is
possible that high speed iron can be observed in the outskirts of the
SN remnant.

Our present simulations do not include radiation transport and the
omission becomes increasingly unrealistic at late times when the
cooling of the ejecta might affect its dynamics. The earlier
fragmentation of ejecta in our simulations may seed the large-scale
inhomogeneity at later times. The magnetar is also assumed here to
have a constant dipole field strength while some decay would
not be surprising.

\subsubsection{Relevance to ASASSN-151h}

Assuming that the bright transient ASASSN-151h was a supernova, which
is still quite controversial \citep{Bro15,Mil15,Mar15}, it is the brightest 
supernova recorded to this date \citep{Don15}.  One interpretation is a
magnetar-illuminated explosion with a very high initial magnetar
energy, 40 B and a relatively low magnetic field strength, $10^{13} -
10^{14}$ G embedded in a stripped core of 5 - 10 \Msun\
\citep{Met15,Ber16,Dai16,Suk16}. We have shown here that mixing and
breakout are sensitive to the magnetar energy. Mixing is maximal when
the energy input by the magnetar greatly exceeds any other source
driving the explosion, and the magnetar decay time scale is comparable
to the expansion time scale for the star. In this regard, the proposed
models for ASASSN-151h are similar to the 1 ms model calculated here
and mixing and breakout should also be similar. Three signatures of
the model are a strong density inversion inside a shell that contains
most of the ejected mass; extensive mixing; and magnetar wind and
radiation breakout. We have not calculated the time-dependent spectrum
of our models, but our results suggest that the spectrum of a 2D model
will differ appreciably from a 1D model. Mixing will result in
composition inversions. The heavy elements will not mostly be in a
shell traveling at a single speed, giving rise to ``boxy'',
``flat-topped'' spectral lines \citep[e.g.][]{Hof04}.  The mixed heavy
elements would have a high velocity dispersion as shown in
\Fig{vel} and magnetar radiation would also leak out earlier and the
breakout transient predicted by \citet{Met14} would have an earlier
onset.

%%%%%%%%%%%%%%%%%%%%%%%%%%%%%%%%%%%%%%%%%%%%%%%%
% Section 4 - Conclusions                      %
%%%%%%%%%%%%%%%%%%%%%%%%%%%%%%%%%%%%%%%%%%%%%%%%
\section{CONCLUSIONS}
\lSect{conclude}

Previous 1D models for magnetar-powered supernovae could not properly
model the fluid instabilities and mixing that necessarily occur when
an energy that is not trivial compared with the kinetic energy of the
ejecta is deposited in a small amount of deeply situated matter. They
instead produced an unphysical density spike that is smeared over a
broader range of radii and mixed in a more realistic 2D hydro
simulation. The mixed region corresponds, approximately, to the mass
of the inner ejecta that had an initial kinetic energy equal to the
energy deposited by the magnetar. If the magnetar energy exceeds the
initial kinetic energy of the entire supernova, breakout will occur on
a time scale given by the time required to roughly double the supernova
energy. After breakout, about $30\%$ of the deposited energy in the
models studied goes into further accelerating the ejecta. Most of the
rest, i.e., that part not further adiabatically degraded, should
appear as light. Assuming a canonical initial supernova energy
(without magnetar input) of $1 \times 10^{51}$ erg, instabilities and
mixing will be a dominant feature when the initial magnetar period is
less than 3 ms, but a less energetic magnetar or radioactivity could
still appreciably alter the spectrum and supernova remnant
morphology. The resulting mixing transform the supernova ejecta into
filamentary structures whose morphology resembles the Crab Nebula.
While our calculations did not include radiation transport, the
filamentary structure found here may be even more enhanced by cooling
or radiative RT instabilities \citep{Kru09, Jia13, Tsa15}.

In a case where the magnetar energy deposited was 0.5 B (out of a
total available 0.8 B) in a 1.2 B explosion, breakout was marginal.
In a more energetic case where 4.8 B (out of an available 20 B) was
deposited, the supernova was shattered and mixing was extensive. This
mixing would have major implications for the color and spectrum of the
supernova and for the morphology of its remnant, and might be
distinguishable from, e.g., circumstellar interaction \citep{Che14b}.
One massive shell impacting another of comparable or lesser mass would
probably lead to less mixing than exploding a star with a bubble of
radiation.

The present calculations are for bare CO cores, chiefly as a matter of
computational efficiency. A larger star would have required a larger
grid, more levels of AMR, and taken longer to run. Many SLSNe are Type
I, however, and stars with extended envelopes and mass loss may be
more likely to brake their cores so that slower magnetar with weaker
fields\citep{Dun92}, are produced \citep{Heg05,Woo06}. Were our cores
to be embedded in low density red or blue supergiants, much more
mixing and fragmentation is expected since the already clumpy ejecta
would seed additional instabilities in the reverse shock. While there
are many ways to mix a supernova, it could be that the extreme mixing
in magnetar-powered supernovae is a diagnostic for their late time
energy input.

The models calculated here should be generally characteristic of other
situations in supernovae where the nonlinear thin shell instability
(NTSI) plays an important role that is events where an enduring
central energy source piles up matter in a dense shell.  An off-center
density maximum necessary has a region where the density decreases
with radius and accelerating that inverted density will result in
mixing. If the energy driving the compression is comparable kinetic
energy of the dense shell, extensive mixing and fragmentation will
occur.  Other examples besides magnetar winds are the decay of
radioactivity, neutrino-driven winds, and colliding shells.  The
energy from the decay of 0.1 \Msun \ of $^{56}$Ni and $^{56}$Co will
release $1.9 \times 10^{49}$ erg, which is comparable to the kinetic
energy in the inner 2 \Msun \ of a typical 15 \Msun
\ supernova. Mixing is likely to occur in at least that volume.

The Crab Nebula, which many of our 2D figures qualitatively resemble,
is believed to have been the low-energy explosion of a star near 10
\Msun \cite[e.g.][]{Smi13} and to have had a low explosion energy
\citep{Yan15}. This is consistent with an explosion powered in part,
or wholly by a neutrino-powered wind \citep{Arc07,Mel15}. The high
velocity wind pushing on an essentially stationary star might be
expected to develop the same instabilities studied here, albeit for
just the first 10 seconds or so. These instabilities might provide the
seeds for subsequent mixing in the magnetar wind.

Colliding shells, such as those produced in pulsational-pair
instability supernovae, are also known to to produce similar density
spikes and 2D mixing like that studied here \citep{Che14a}.  In future
papers, we will use the radiation transport capabilities of the
\CASTRO\ code \citep{Zha13} to better examine these explosions and
provide more realistic observable diagnostics.

\acknowledgements The authors are grateful to the anonymous referee for providing insightful comments.
The authors thank Ann Almgren and Weiqun Zhang for
help with the \CASTRO\ code. We also thank Alexander Heger, Mark
Krumholz, and Dan Kasen for useful discussions. K.C. acknowledges
the support of EACOA Fellowship from the East Asian Core Observatories
Association and the hospitality of Aspen Center for Physics, which is
supported by National Science Foundation grant PHY-1066293. Work at
UCSC has been supported by an IAU-Gruber Fellowship, the DOE HEP
Program (DE-SC0010676) and the NASA Theory Program
(NNX14AH34G). \CASTRO\ was developed through the DOE SciDAC program by
grants DE-AC02-05CH11231, and DE-FC02-09ER41618.  Numerical
simulations are supported by the National Energy Research Scientific
Computing Center (NERSC), and the Center for Computational
Astrophysics (CfCA) at National Astronomical Observatory of Japan
(NAOJ).

% REFERENCES
%%%%%%%%%%%%%%%%%%%%%%%%%%%%%%%%%%%%%%%%%%%%%%%%%%%%%%%%%%%%%%

%%%%%%%%%%%%%%%%%%%%%%%%%%%%%%%%%%%%%%%%%%%%%%%%%%%%%%%%%%%%%%
\clearpage
\end{document}